\documentclass[conference]{IEEEtran}
\IEEEoverridecommandlockouts
\usepackage{cite}
\usepackage{amsmath,amssymb,amsfonts}
\usepackage{algorithmic}
\usepackage{graphicx}
\usepackage{textcomp}
\usepackage{xcolor}

\usepackage{booktabs}
\usepackage{makecell}
\usepackage{subfigure}
\usepackage{algorithm}
\usepackage{algorithmic}
\usepackage[switch]{lineno}
\usepackage{color}

\usepackage{etoolbox}
\makeatletter
\patchcmd{\@makecaption}
  {\scshape}
  {}
  {}
  {}
\makeatother
\def\tablename{Table}

\newcommand{\eg}{{\em e.g.,~}}

\def\BibTeX{{\rm B\kern-.05em{\sc i\kern-.025em b}\kern-.08em
    T\kern-.1667em\lower.7ex\hbox{E}\kern-.125emX}}
\begin{document}

\title{Portfolio-Based Incentive Mechanism Design for Cross-Device Federated Learning}

\author{
  Jiaxi Yang\textsuperscript{1},
  Sheng Cao\textsuperscript{1},
  Cuifang Zhao\textsuperscript{2}, 
  Weina Niu\textsuperscript{1},
  Li-Chuan Tsai \textsuperscript{3} 
  \thanks{Correspondence to: Li-Chuan Tsai $<$1201500112@jxufe.edu.cn$>$} \\
  \textsuperscript{1}University of Electronic Science and Technology of China \\
  \textsuperscript{2}National Taiwan University  \\
  \textsuperscript{3}Jiangxi University of Finance and Economics\\

}

\maketitle

\begin{abstract}
In recent years, there has been a significant increase in attention towards designing incentive mechanisms for federated learning (FL). Tremendous existing studies attempt to design the solutions using various approaches (\eg game theory, reinforcement learning) under different settings. Yet the design of incentive mechanism could be significantly biased in that clients' performance in many applications is stochastic and hard to estimate. Properly handling this stochasticity motivates this research, as it is not well addressed in pioneering literature. In this paper, we focus on cross-device FL and propose a multi-level FL architecture under the real scenarios. Considering the two properties of clients' situations: uncertainty, correlation, we propose FL Incentive Mechanism based on Portfolio theory (FL-IMP). As far as we are aware, this is the pioneering application of portfolio theory to incentive mechanism design aimed at resolving FL resource allocation problem. In order to more accurately reflect practical FL scenarios, we introduce the Federated Learning Agent-Based Model (FL-ABM) as a means of simulating autonomous clients. FL-ABM enables us to gain a deeper understanding of the factors that influence the system's outcomes. Experimental evaluations of our approach have extensively validated its effectiveness and superior performance in comparison to the benchmark methods.
\end{abstract}


\section{Introduction}
Federated Learning (FL) is a popular distributed machine learning paradigm that enables collaborative model training without requiring data owners to share their private data. The privacy protection advantages of FL have gained significant attention from both academic and industrial fields. However, to obtain high-quality models in FL, more clients are needed for local training, which inevitably consumes resources such as computation and communication costs. Additionally, recent advanced attack approaches have increased the risk of privacy leakage for clients in FL. As a result, no client is willing to participate in FL without reasonable compensation. Therefore, designing an incentive mechanism for FL is crucial in light of these considerations.

In previous studies, researchers have explored designing incentive mechanisms to induce clients' participation based on simply assumptions on clients' capacities. For example, Zhang et al.~\cite{zhang2021incentive} leverage historical clients' performance to estimate their current capacities and reputation, and design a reverse auction-based incentive mechanism. A reputation-based clients selection scheme is also combined with contract theory for FL incentive mechanism design~\cite{kang2019incentive}. Other studies consider the relationship between clients' hardware and their computation capacity or assume a deterministic capacity in problem settings~\cite{sarikaya2019motivating, khan2020federated, pandey2020crowdsourcing, zhan2020incentive}. 
However, none of these studies fully address the challenge of accurately estimating clients' capacities, such as their communication and computation abilities, in each round. One major challenge in estimating clients' capacities is that in cross-device FL, clients are often Internet of Things (IoT) devices with limited computation and communication abilities. The status of these IoTs can be highly dependent on factors such as battery level and communication channels, which can introduce uncertainty and instability over time. Another challenge in accurately estimating clients' capacities is that clients' situations may be correlated. This is because different clients may share similar constraints, which may come from their participating activities, exposed environments, and their self-physical constraint. As a result, estimating clients' capacities is often biased, which can lead to suboptimal choices when using existing incentive mechanisms in FL. Resolving these challenges is the core motivation of our work.



In this paper, we consider a universal cross-device FL, which contains wireless communication infrastructures and numerous IoTs. Then we propose a multi-level cross-device FL architecture with the following contributions. To allocate rewards under the limited budget and obtain the total utility optimization, we take the stochastic status of clients into consideration. By analyzing clients' historical performance and their correlation property in our FL architecture, portfolio theory is firstly leveraged to address this issue. The main contributions of our work can be summarized in the following points:
\begin{itemize}
    \item Our proposed multi-level cross-device FL architecture is tailored to the demands of a universal cross-device FL system featuring wireless communication infrastructures and a multitude of IoTs. The architecture is crafted to be highly adaptable and practical for real-world implementation.
    \item Through our analysis of the uncertainty and correlation properties of clients in cross-device FL, we investigate the challenges related to accurately estimating their capacities in light of these factors. As pioneers in this domain, we firstly propose the application of portfolio theory to tackle the capacity estimation challenges in cross-device FL and design an effective incentive mechanism for it.
    \item To better capture the practical FL scenarios, we propose the Federated Learning Agent-based Model (FL-ABM) to simulate autonomous clients of FL. Our extensive experimental results demonstrate that our approach is significantly more effective and outperforms the benchmark methods.
\end{itemize}

\section{Related Work}

\begin{table*}[!htbp]
\centering
\caption{The comparison of some notable works of FL incentive mechanism: Incentive Compatibility (IC), Individual Rationality (IR), Pareto Efficiency (PE)}
\begin{tabular}{cclllc}
\hline
Reference & Main Technology & IC & IR & PE & Description \\ 
\hline
 \cite{sarikaya2019motivating} & Stackelberg Game &  & \checkmark &    & clients determine the CPU power they use to participate            \\ 
 \cite{khan2020federated} & Stackelberg Game &    & \checkmark    &    &  clients determine the optimal CPU-frequency strategy           \\ 
\cite{pandey2020crowdsourcing} & Stackelberg Game &    &\checkmark    &    & clients determine the number of iterations with the fixed communication time \\ 
\cite{zhan2020incentive} & DRL &    &\checkmark &    &   clients determine the computation capability          \\ 
\cite{zhang2021incentive} & Reputation Mechanism, Auction Theory & &\checkmark &    & selecting clients based on their historical reputation  \\ 
\cite{kang2019incentive} & Reputation Mechanism, Contract Theory & \checkmark &\checkmark &\checkmark &  server contract items to clients and select clients by their historical performance\\ 
\cite{zhan2020learning} & DRL & & \checkmark&    &  \begin{tabular}[c]{@{}c@{}}Considering the fixed communication and computation cost, \\clients determine their participants level with the total reward\end{tabular} \\ 
\cite{zeng2020fmore} & Auction Theory & \checkmark &\checkmark &\checkmark & clients receive a bid ask from server under the requirement of resources \\
\cite{jiao2020toward} & Auction Theory, DRL & \checkmark& \checkmark &    &  assuming the data owners’ capability can meet the requirement\\
\cite{tang2021incentive} & Public Goods & & \checkmark   & \checkmark & consider the deterministic costs for per processing capacity \\
\cite{ding2021incentive} & Stackelberg Game & \checkmark & \checkmark &    &  
\begin{tabular}[c]{@{}c@{}}The computation time of each worker\\ follows cumulative distribution function (CDF) \end{tabular}
 \\
\hline
\end{tabular}
\label{literatures}
\vspace{-5mm}
\end{table*}


A key prerequisite for effective incentive mechanism design in FL is providing sufficient rewards to induce clients to participate in collaborative training. 
Previous studies in this field focus on addressing two sub-problems: fairness and optimization.
While contribution measurement approaches such as Shapley value and its variations have been widely used to address fairness problems in previous work~\cite{sim2020collaborative, liu2022gtg}, they are less relevant for the optimizations, and particularly not related to biased sub-optimality issues mentioned in this paper. 
To address the optimization problem, pioneers apply game theory-based methods such as auction theory, Stackelberg game, and contract theory for incentive mechanism design~\cite{zeng2020fmore, yang2023federated, sarikaya2019motivating, kang2019incentive}. Deep reinforcement learning, a common means to solve dynamic problems in game theory, is also leveraged to compute Nash equilibrium and social welfare maximum~\cite{zhan2020learning, gu2020multiagent, zhan2020incentive, jiao2020toward}. 

Zeng et al. propose FMore, in which the clients submit their bids with promised capacities to the aggregation server~\cite{zeng2020fmore}. 
In their paper~\cite{jiao2020toward}, Jiao et al. do not take into account the stochastic property of clients' capacities, and propose an incentive mechanism for maximizing social welfare using deep reinforcement learning.
In~\cite{zhan2020incentive}, clients deterministically decide their participation level based on  per unit costs of communication and computation under given the total reward. The studies by Kang et al.~\cite{kang2019incentive} and Zhang et al.~\cite{zhang2021incentive} both employ a reputation mechanism for estimating clients' capacities and selecting nodes, by exploiting their historical performance. Zhan et al.~\cite{zhan2020learning} propose an incentive mechanism, in which the clients automatically learn the best strategy based on deep reinforcement learning. In other settings, clients are able to determine their computation capacity to optimize their utility function~\cite{sarikaya2019motivating, khan2020federated, pandey2020crowdsourcing}. However, all these works do not consider the uncertainty of clients' capacities over time, which can be obviously observed in Table~\ref{literatures}.

\begin{figure}[t]
\setlength{\abovecaptionskip}{-0.10cm}
	\centering{\includegraphics[scale = 0.3]{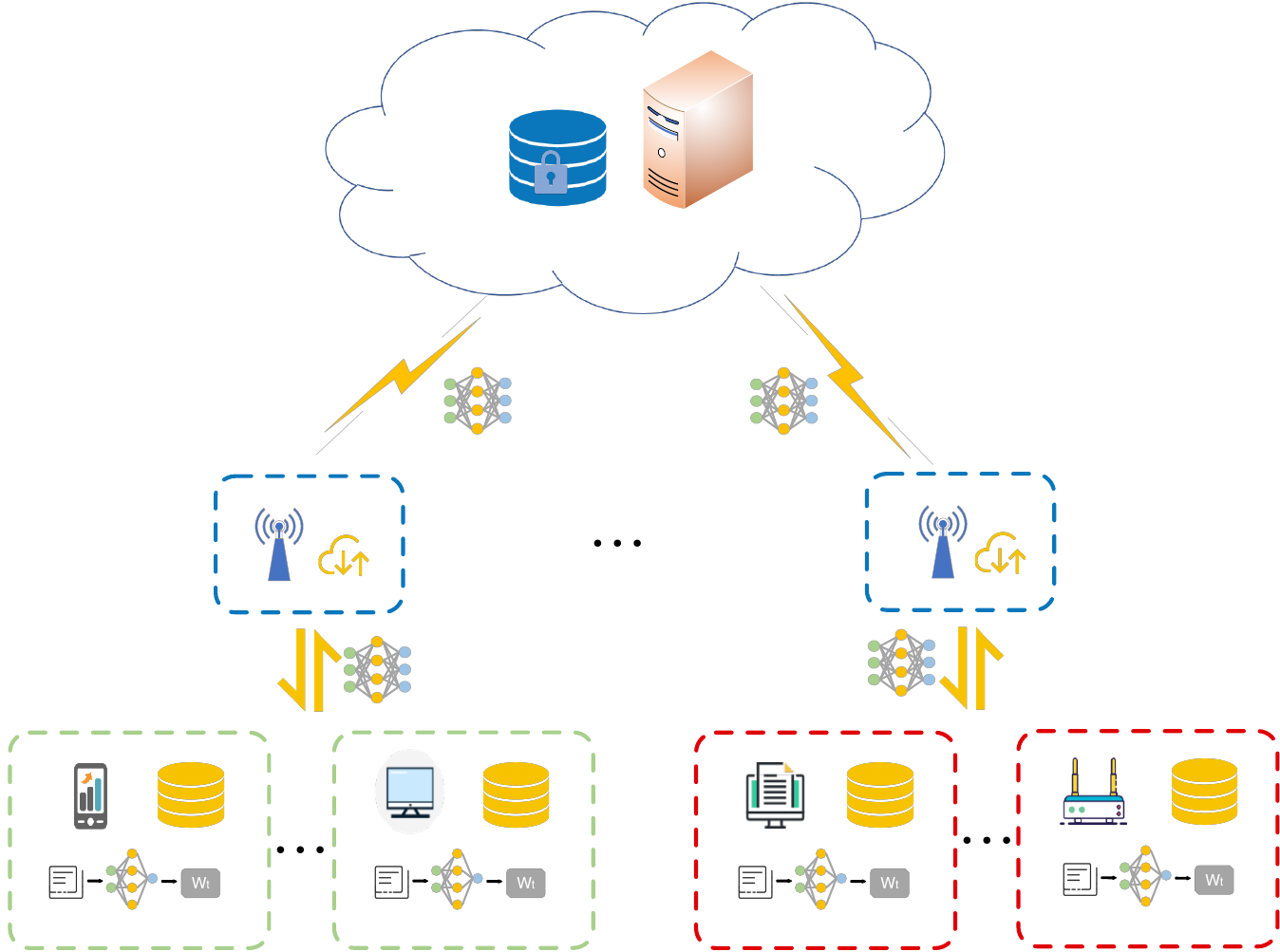}}
	\caption{Overview of Multi-level Federated Learning Architecture. The three layers from top to bottom consist of aggregation servers (MBSs), edge computing devices (EBSs), and clients (IoTs).}
	\label{architecture}
\end{figure}

\section{Multi-level Federated Learning Architecture}
In vanilla FL model training process, the aggregation server coordinates with different clients to optimize the objectives collaboratively. Here, the objective function of the optimization is formulated as:
\begin{gather}
    \min_{w \in \mathbb{R}^{d}} \mathcal{F}(\textbf{w}) := \frac{1}{N} \sum_{n=1}^{N} f_{n}(\textbf{w}), \\
    f_{n}(\textbf{w}) \triangleq \mathbb{E}_{(x,y)\sim \mathcal{X}}[f_{n}(\textbf{w}; x,y)],
    \label{ObjectiveFL}
\end{gather}

where $\textbf{w}$ is the parameters of the global model and $f_{n}(\textbf{w})$ is the local loss function. Based on the number of clients $N$, FL can be broadly divided into two categories: cross-device FL and cross-silo FL. The fundamental disparity between cross-silo settings and cross-device FL is that the former involves learning across a limited number of organizations that possess ample data, while the latter involves Internet of Things (IoT) clients, which can number in the millions. Consequently, the clients in cross-device FL are much more unstable due to varying factors’ status such as battery level, communication channel, and other limited resources. This results in unpredictable fluctuations in the number of time steps taken by IoTs. 

This paper focuses on cross-device FL and proposes a multi-level federated learning architecture comprising IoTs, Edge Base Stations (EBSs), and Macro Base Stations (MBSs), as depicted in Fig.~\ref{architecture}. The FL aggregation servers in this architecture can be high-performance MBSs with abundant communication and computation resources, while the clients can be mobile devices with limited communication and computation resources. Notably, most of these mobile devices' conditions are unstable in real-time due to the aforementioned reasons. As for EBSs, their computation and communication capabilities are intermediate between those of MBSs and IoTs.

During the training process, MBSs function as aggregation servers and initialize the global model, which they then send model parameters to EBSs. Subsequently, EBSs regulate a group of IoTs and transfer these parameters to them. Like in conventional FL, these IoTs use their own data and local resources to update parameters and send them back to the transfer station, EBSs. Finally, MBSs are employed to perform an aggregation algorithm, such as FedAvg, on the updated model parameters from the EBSs.

\section{Problem Formulation}
We denote the reward budget of the aggregation server as $B$, while the capacity of clients is affected by unstable communication and computation resources. Furthermore, the price of a one unit of resource for the $i$-th client is designated by $c_i$.
Let $\textbf{r} = \{ r_{1}, r_{2}, ..., r_{n} \}$ be the expected benefit returns of clients' contribution in FL task, and $\bar{r} = \mathbb{E}(r)$. $R_{p}$ is the estimated total benefit return for global model. In particular, the expected benefit return $r_i$ can be measured by the performance improvement, which is defined in equation (\ref{r}) and (\ref{R}). And we use test accuracy of the machine learning model to be the performance metrics.
\begin{gather}
    r_i = acc_{i}^{t} - acc_{i}^{t-1}.
    \label{r}
    \\
    R_{p} = acc_{global}^{t} - acc_{global}^{t-1}. 
    \label{R}
\end{gather}
However, in the FL settings, two critical concerns regarding clients necessitate our attention: uncertainty and correlation. 1) Throughout the process of model training, the circumstances of clients, such as their computation and communication capacities, may fluctuate over time and are difficult to precisely estimate. 2) IoTs in the same region are managed by a EBS, which leads to communicaton status of IoTs covary to some extent in each time step $t$. Due to the resource constraints of IoTs, the situations of their computation and communication capacities are tied positively or negatively. To properly resolve proceeding concerns, we borrow ideas of portfolio theory to handle one branch of FL incentive mechanism problem: resource allocation~\cite{zeng2021comprehensive}, with formulations detailed in next section.
\begin{algorithm}[b]
    \caption{Portfolio-based Incentive Mechanism for FL}
    \label{alg:algorithm}
    \textbf{Input}: The total reward $B$, historical return $\textbf{r}_{history}$\\
    \textbf{Output}: Global model $\mathcal{M}_{global}$ 
    \begin{algorithmic}[1] 
        \STATE Initialize global model model $\mathcal{M}_{global}$
        \STATE /* Allocation strategy for clients by FL-IMP */
        \STATE $w_{1},...,w_{n} \leftarrow \textit{PortfolioAllocating}(\left\{c_{1},...,c_{i}\right\}, \textbf{r}_{history})$
        \FOR{each round $t=1,2,..., T$}
            
            \FOR{each client $i$}
                \STATE /* Compute the local model update */
                \STATE $\mathcal{M}_{i,local} \leftarrow \textit{LocalTraining}(i, \mathcal{M}_{global}, w_{i})$
            \ENDFOR
            \STATE $\mathcal{M}_{global} = \frac{1}{n} \sum_{i=0}^{n} \mathcal{M}_{i,local}$ 
        \ENDFOR
        \STATE \textbf{return} $\mathcal{M}_{global}$
    \end{algorithmic}
\end{algorithm}

\section{Portfolio based Incentive Mechanism of FL}
The task of resource allocation of FL incentive mechanism design is to greatly improve the model performance and maximize the total benefit return $R_{p}$ under the budget $B$ of the aggregation server. And the design of it bears similarities to the investment problem in financial markets as described by portfolio theory. The situation that server allocates rewards on clients is analogous to investors’ choices on stocks, both face how to optimize allocations under stochastic environment.

An immediate popped out solution is to conduct the following seemingly effective strategy:

\begin{equation}
    \mathbb{E}[R_{p}] = \sum^{n}_{i=1}w_{i} \cdot \mathbb{E}[r_{i}] = \mathbf{w} \overline{\mathbf{r}}^{\prime},
    \label{weighted}
\end{equation}

a variable $w_{i}$, where $\sum_{i=1}^{n}w_{i}=1$, is assigned to each of the compensation for clients for differentiating the allocated reward. It can be easily shown that when $w_1 > w_2... >w_{n}$, the maximum value of $R_n$ gives the amount of reward allocation order $r_1 > r_2 ... > r_n$.
This maximization follows immediately by setting $w_n$ positively correlatively with $r_n$.

However, the performance improvement of global model $R_n$ can not be calculated with certainty. It relies on the estimations of the performance improvement of each local model.
As in the previous discussion, uncertainy naturally exists in practical FL applications. To address such uncertainty, we make a probability statement about model performance improvement, assuming the model performance improvement is random variables and have their own probability distributions. Their joint distribution is summarized by means and covariance matrix. The off-diagonal element $\sigma_{i,j}$ in the matrix indicates the covariance of the model performance improvement between the client $i$ and client $j$; the diagonal element $\sigma_{i, i}$ is the variance of the performance improvement of local model, which indicates the dispersion from the mean $E[r_i]$. $\sigma_{i} = \sqrt{\sigma_{i,i}}$ is the standard deviation, and $\rho_{i,j} = \frac{\sigma_{i,j}}{\rho_{i}\rho_{j}}$ is the correlation coefficient of two clients. The variance of the portfolio is given by:
\begin{equation}
\sigma^{2}\left(R_{p}\right)=\mathbf{w}^{\prime}\mathbf{Vw}=2\sum_{i=1}^{n-1}\sum_{j=I+1}^{n}w_{i}w_{j}\sigma_{ij}+\sum_{i=1}^{n}w_{i}^{2}\sigma_{i}^{2}.
\label{eq:portfolio_variance}
\end{equation}









Equation (\ref{eq:portfolio_variance}) indicates that the server's
risk depends not only on any given client's risk, but also on correlation
component. From (\ref{eq:portfolio_variance}), we can also derive:
\begin{equation}
    \frac{\partial\sigma^{2}\left(R_{p}\right)}{\partial w_{i}}=2\sum_{j=1}^{n}w_{j}\sigma_{ij}.
\end{equation}

Furthermore, if we express the correlation of client $i$ and portfolio
$P$ as $\sigma_{i,P}$, then it's straightforward to show that
\begin{equation}
    \begin{aligned}        \sum_{j=1}^{n}w_{j} \sigma_{ij}=&\sum_{j=1}^{n}w_{j}Cov\left(r_{i},r_{j}\right)=Cov\left(r_{i},\sum_{j=1}^{n}w_{j}r_{j}\right)\\
    &=Cov\left(r_{i},R_{p}\right)=\sigma_{iP}.
    \end{aligned}
\end{equation}

Thus, 
\begin{equation}
    \frac{\partial\sigma^{2}\left(R_{p}\right)}{\partial w_{i}}=2\sigma_{iP}.
\end{equation}

Above result has an important implication that the marginal contribution
of client $i$ to the portfolio risk is twice its correlation with
the portfolio.

In our work, we leverage Markowitz Portfolio Theory~\cite{Markowitz} and find a straightforward way to describe the effectiveness of the resource allocation strategy in FL. Following the Markowitz approach, we have to determine the set of clients, which minimize the variance for given expected performance improvement of global model $\mathbb{E}\left[R_{p}\right]$. This leads to the following quadratic optimization problem:

\begin{equation}
    \begin{gathered}\min_{\mathbf{w}}\mathbf{w}^{\prime}\mathbf{V}\mathbf{w},\\
    \text{ with }\mathbf{w}^{\prime}\overline{\mathbf{r}}=\mathbb{E}\left[R_{P}\right],
    \mathbf{w}^{\prime}\mathbf{e}=1.
    \end{gathered}
\end{equation}

where $\mathbf{V}=\left[\sigma_{ij}\right]_{1\le i,j\le n}$ is the $n\times n$ variance-covariance matrix of local model improvements, and is assumed to be invertible, and $\mathbf{e}$ is the vector of ones.
To tackle this problem, consider the following Lagrangian functional:
\begin{equation}
    \begin{aligned}       L(\mathbf{w},\lambda,\delta)=&\mathbf{w}^{\prime}\mathbf{V}\mathbf{w}+\lambda_{1}\left(\mathbb{E}\left[R_{P}\right]-\mathbf{w}^{\prime}\overline{\mathbf{r}}\right)\\ 
    &+\lambda_{2}\left(1-\mathbf{w}^{\prime}\mathbf{e}\right).
    \end{aligned}
\end{equation}

where $\lambda_{1}$ and $\lambda_{2}$ are constant Lagrangina multipliers.
This problem is then equivalent to :

\begin{equation}
    \begin{aligned}
    \min_{ \{\mathbf{w},\lambda_{1},\lambda_{2}\} } L(\mathbf{w},\lambda_{1},\lambda_{2}) = & \mathbf{w}^{\prime}\mathbf{V}\mathbf{w} +\lambda_{1} \left( \mathbb{E}\left[R_{P}\right] - \mathbf{w}^{\prime} \overline{\mathbf{r}}  \right) \\
    & +\lambda_{2}\left(1-\mathbf{w}^{\prime}\mathbf{e}\right).
    \end{aligned}
\end{equation}

The first order conditions are:

\[
\begin{aligned} & \frac{\partial L(\mathbf{w},\lambda_{1},\lambda_{2})}{\partial\mathbf{w}}=2\mathbf{V}\mathbf{w}-\lambda_{1}\overline{\mathbf{r}}-\lambda_{2}\mathbf{e}=0,\\
 & \frac{\partial L(\mathbf{w},\lambda_{1},\lambda_{2})}{\partial\lambda_{1}}=\mathbb{E}\left[R_{P}\right]-\mathbf{w}^{\prime}\overline{\mathbf{r}}=0,\\
 & \frac{\partial L(\mathbf{w},\lambda_{1},\lambda_{2})}{\partial\lambda_{2}}=1-\mathbf{w}^{\prime}\mathbf{e}=0.
\end{aligned}
\]

As $\mathbf{V}$ is assumed to be invertible, the first-order conditions
are both necessary and sufficient for the following unique solution:
\begin{equation}
    \begin{aligned}
        \mathbf{w}=&\frac{1}{\theta_{4}}\left(\theta_{2}\mathbf{V}^{-1}\mathbf{e}-\theta_{1}\mathbf{V}^{-1}\overline{\mathbf{r}}\right)+\\
        &\mathbb{E}\left[R_{P}\right]\frac{1}{\theta_{4}}\left(\theta_{3}\mathbf{V}^{-1}\overline{\mathbf{r}}-\theta_{1}\mathbf{V}^{-1}\mathbf{e}\right). \\ 
    \end{aligned}
    \label{w}
\end{equation}

where $\theta_{1}=\mathbf{e}^{\prime}\mathbf{V}^{-1}\overline{\mathbf{r}},\theta_{2}=\overline{\mathbf{r}}^{\prime}\mathbf{V}^{-1}\overline{\mathbf{r}},\theta_{3}=\mathbf{e}^{\prime}\mathbf{V}^{-1}\mathbf{e}\text{ and }\theta_{4}=\theta_{2}\theta_{3}-\theta_{1}^{2}$. And then, (\ref{w}) can be further derived to obtain

\begin{equation}
\frac{\sigma^{2}\left(R_{p}\right)}{\theta_{3}}-\frac{\left[\mathbb{E}\left(R_{p}\right)-\theta_{1}/\theta_{3}\right]^{2}}{\theta_{4}/\theta_{3}^{2}}=1.
\label{eq:relation}
\end{equation}

(\ref{eq:relation}) is a hyperbola on the $\left(\sigma\left(R_{p}\right),\mathbb{E}\left(R_{p}\right)\right)$
plane whose arc constrains the set of all possible reward allocation strategies for clients. As
any rational agent would prefer more than less, the upper portion
of this arc is referred to as \textit{efficient frontier}~\cite{portfolioSelection}. The asymptotes
are determined by 
\begin{equation}
   \mathbb{E}\left(R_{P}\right)=\frac{\theta_{1}}{\theta_{3}}\pm\frac{\theta_{4}/\theta_{3}^{2}}{1/\theta_{3}}\sigma\left(R_{P}\right). 
   \label{expectedReturn}
\end{equation}

At a given performance improvement of FL aggregation server $\mathbb{E}\left(R_{p}\right)$,
(\ref{expectedReturn}) implicitly suggests the lowest possible variance.
More importantly, the minimum variance lies at the vertex of this
hyperbola, with value equal to $\sqrt{1/\theta_{3}}$, and the corresponding performance improvement is $\theta_{1}/\theta_{3}$.

\begin{figure*}[ht]
\setlength{\abovecaptionskip}{-0.20cm}
	\centering{\includegraphics[scale = 0.48]{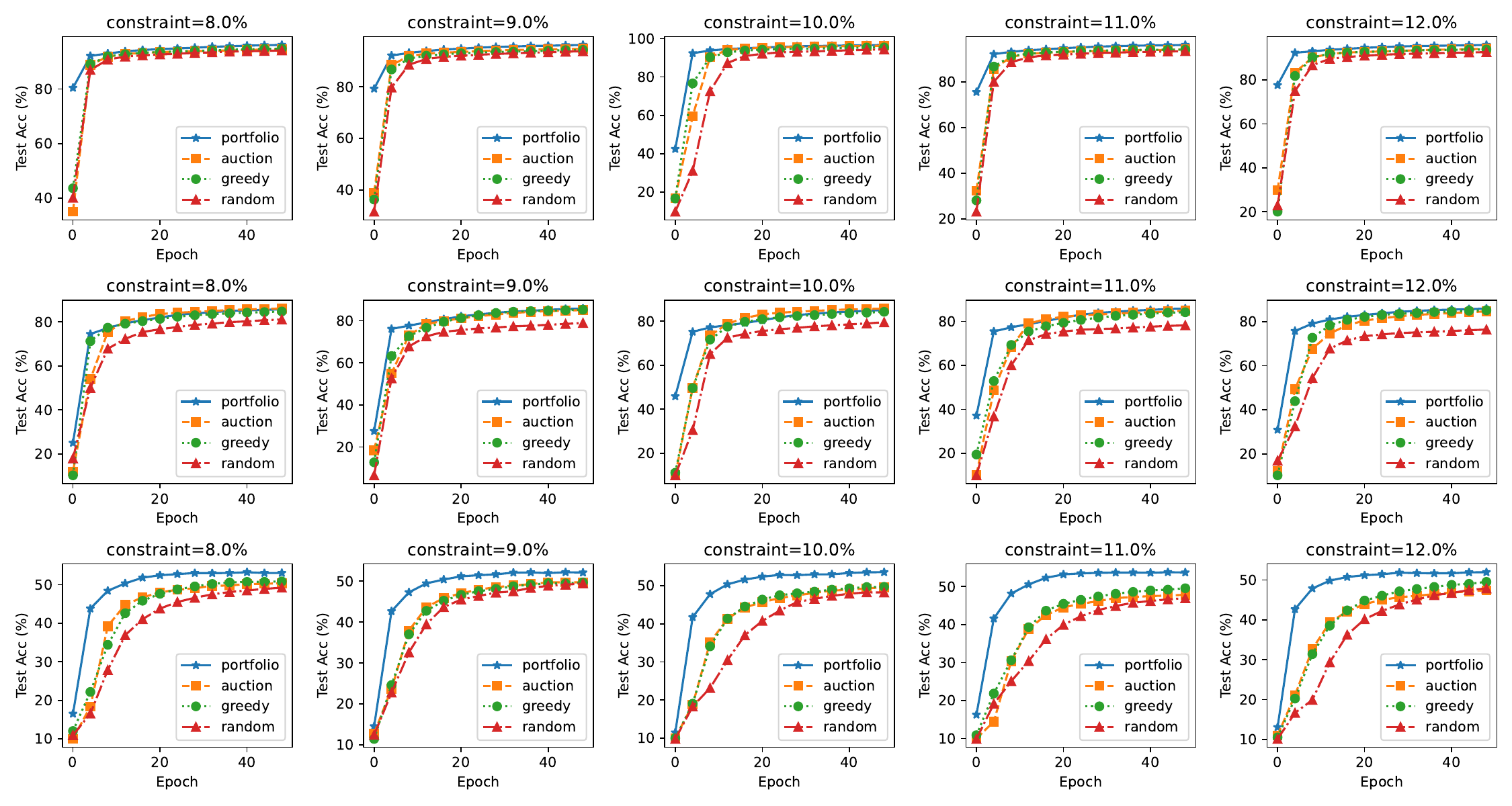}}
	\caption{Test accuracy during the training process on MNIST, Fashion-MNIST and CIFAR-10, arranged in ascending order from top to bottom}
	\label{accuracy}
 \vspace{-2mm}
\end{figure*}

\begin{table}[b]
\vspace{-5mm}
\caption{Types of agents, their attributions and actions}
\begin{tabular}{lllll}
\hline
\multicolumn{5}{c}{A1: IoTs}                                                                                                                                                                                   \\ \hline
\multicolumn{1}{c}{Description} & \multicolumn{4}{l}{\begin{tabular}[c]{@{}l@{}}Its dynamic communication and computation capacities \\ varies in each iteration and may be associated with A2\end{tabular}} \\ 
Attributes                      & \multicolumn{4}{l}{\begin{tabular}[c]{@{}l@{}}FL Tasks, Lower Clients (EBSs, A2), Communication \\ Resources, Computation Resources\end{tabular}}                            \\ 
Actions                         & \multicolumn{4}{l}{Participate in a new FL task, Leave a FL task, Break down}                                                                                                \\ \hline
\multicolumn{5}{c}{A2: EBSs}                                                                                                                                                                                   \\ \hline
Description                     & \multicolumn{4}{l}{\begin{tabular}[c]{@{}l@{}}The leader manages IoTs (A1) in a region and interact\\ with MBSs (A3).\end{tabular}}                                          \\ 
Attributes                      & \multicolumn{4}{l}{Managers of the group (A1), Upper clients of  MBSs (A3)}                                                                                                  \\ 
Actions                         & \multicolumn{4}{l}{\begin{tabular}[c]{@{}l@{}}Manage a new FL task, Leave a FL task, Communicate with\\ A3, Communicate with conducted group A1\end{tabular}}                \\ \hline
\multicolumn{5}{c}{A3: MBSs}                                                                                                                                                                                   \\ \hline
Description                     & \multicolumn{4}{l}{\begin{tabular}[c]{@{}l@{}}The aggregation servers of the FL tasks which is responsible\\ to coordinate FL training\end{tabular}}                         \\ 
Attributes                      & \multicolumn{4}{l}{FL task,  Managers of the upper clients group (A2)}                                                                                                       \\ 
Actions                         & \multicolumn{4}{l}{Communicate with A2}                                                                                                                                      \\ \hline

\end{tabular}\\

\label{abm}
\end{table}

\section{Experiments}
\subsection{Experiment Settings}
In this work, extensive experiments are conducted to evaluate the performance of the proposed resource allocation of FL incentive mechanism. 
\\
\noindent \textbf{Simulation:}
In order to better simulate the practical scenarios of federated learning and verify the well-performance of our approach, we firstly leverage FL-ABM in our experiments~\cite{van1998agent}. Agent-Based Simulations (ABS) are a good choice to simulate such systems, due to their effecitveness of implementation and accurate results when compared with real data~\cite{sichman2005multi}. We develop FL-ABM to simulate the dynamic uncertainty of the clients’ situations and also their correlation. The proposed FL-ABM aims to emulate a practical scenario of multi-level FL, consisting of agents that represent IoTs, EBSs and MBSs, each one with specific attributes and behaviors. The settings of FL-ABM is presented as in TABLE~\ref{abm}. In the scenario, there are 15 EBSs managing variable groups of lower clients, and each of the groups contains different number of IoT clients, from 10 to 15. The uncertainty of per unit capacity for each client varies up to $15\%$ in each round. And the unit of the clients' initial computation capability varies from 3 to 7. 
\\
\noindent \textbf{Datasets and Models:}
In this experiments, we train three different models using three different datasets: Digital MNIST~\cite{lecun1998gradient}, Fashion MNIST~\cite{xiao2017fashion}, and CIFAR-10~\cite{krizhevsky2009learning}. Specifically, we will use a Multi-Layer Perceptron (MLP) for Digital MNIST, a Convolutional Neural Network (CNN) for Fashion MNIST and CIFAR-10. We employ an MLP architecture with multiple layers of nodes to train the MNIST dataset, whereas a CNN architecture with two convolutional layers and two fully connected layers is used for Fashion MNIST. Additionally, we also utilize CIFAR-10 dataset, which is trained using a CNN comprising of two convolutional layers with max-pooling, followed by three fully connected layers.

\subsection{Benchmarks}
To evaluate the effectiveness of our proposed approach for selecting clients in FL, we compare it against three baseline methods: the auction-based method, the random-based method, and a greedy-based method.

\begin{itemize}
    \item \textit{\textbf{Random-based Approach}}
    The random-based method selects clients randomly in FL community, without considering their past performance. 
    \item \textit{\textbf{Greedy-based Approach}}
    The greedy-based method selects the client that performed the best in the last FL iterations, constrained by the upper investment limit for each client.
    \item \textit{\textbf{Auction-based Approach}} 
    In this baseline method, we choose the winning client based on its average cost-effectiveness in past FL rounds, with same aforementioned constraint.
\end{itemize}


\vspace{-4mm}
\begin{table}[h]
\setlength{\abovecaptionskip}{-0.10cm}
\caption{Utility with different budget.}
\centering
  \resizebox{0.9\columnwidth}{!}{%
\begin{tabular}{cccc}
\toprule
\textbf{Budget} & \textbf{MNIST}  & \textbf{Fashion-MNIST}   & \textbf{CIFAR-10}   \\ \hline
$B=400$ & $9.234 \pm 0.014$ & $8.133 \pm 0.013$ & $4.598 \pm 0.025$  \\
$B=500$ & $9.277 \pm 0.012$ & $8.194 \pm 0.009$ & $4.652 \pm 0.034$  \\
$B=600$ & $9.320 \pm 0.012$ & $8.236 \pm 0.001$ & $4.705 \pm 0.021$ \\
$B=700$ & $9.346 \pm 0.008$ & $8.276 \pm 0.017$ &  $4.760 \pm 0.052$ \\ 
$B=800$ & $9.373 \pm 0.014$ & $8.310 \pm 0.016$ & $4.808 \pm 0.034$ \\
$B=900$ & $9.389 \pm 0.013$ & $8.330 \pm 0.015$ & $4.829 \pm 0.012$ \\\hline

\end{tabular}}

\label{experiment:budget}
\vspace{-5mm}
\end{table}

\subsection{Experiment Results}
In order to conduct a more thorough examination of the experimental outcomes, we individually evaluate the performance of our approach and the three baseline methods on each of the three datasets. This evaluation enables us to ascertain whether our approach demonstrates consistent excellence across all datasets.

We first conduct experiments to observe the performance difference between our approach and the baseline methods. The findings, presented in Fig.~\ref{accuracy}, reveal that our proposed approach outperforms the three baseline approaches in terms of faster convergence rate and superior performance across all three datasets. Notably, the random-based approach exhibits the poorest performance, whereas the auction-based and greedy-based methods exhibit performance levels between our approach and the random-based method. These results remain consistent under varying budget constraint levels ranging from $8\%$ to $12\%$. Therefore, we can confidently assert that our approach's superiority is not merely due to chance, but rather a significant improvement over the baseline methods.

Besides, we examine how the performance of each method is affected by the budget. We accomplish this by plotting average $R_{t}$ of the beginning 10 iterations for various budgets and by analyzing the trend in performance as the budget increases from 400 to 900. As displayed in Table~\ref{experiment:budget} for the three different datasets, the average utility return increases with the budget's rise. Overall, a comprehensive analysis of our findings can furnish valuable verification of the efficacy and efficiency of our approach.

\begin{figure}
\setlength{\abovecaptionskip}{-0.10cm}
  \centering
  \begin{subfigure}
  \centering{\includegraphics[scale=0.33]{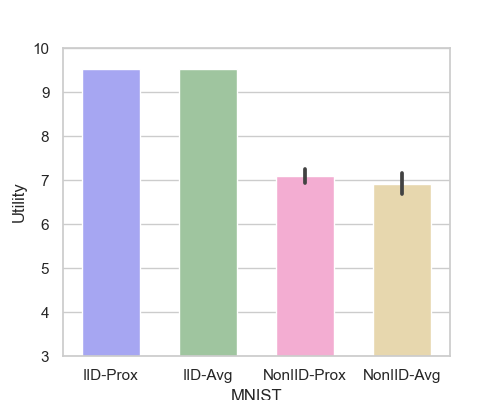}}
  \label{fig:subfig1}
  \end{subfigure}
  \begin{subfigure}
  \centering{\includegraphics[scale=0.33]{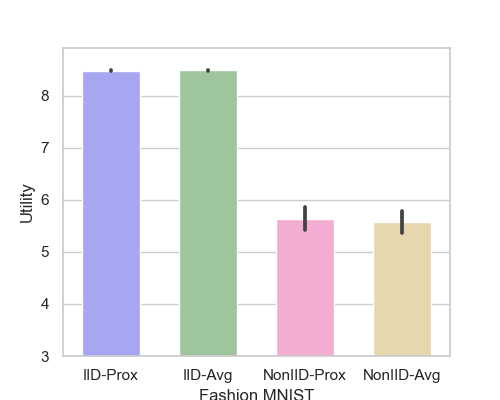}}
  \label{fig:subfig1}
  \end{subfigure}
  
  \caption{The utility with FedAvg and FedProx on IID and Non-IID Dataset.}
  \label{avgAlgorithm}
  \vspace{-5mm}
\end{figure}

To further validate the universality of our approach, we examine it with two different aggregation algorithms: FedAvg, FedProx~\cite{li2020federated}. The datasets used in this experiment included MNIST and Fashion MNIST, which were evaluated under both Independent and Identically Distributed (IID) and Non-IID data partition settings. The results in Fig.~\ref{avgAlgorithm} illustrate that our approach with different aggregation algorithms performed similarly, while FedProx performed slightly better under Non-IID settings.  

\section{Conclusion}
In this paper, we design an incentive mechanism for resource allocation among different heterogeneity clients in cross-device FL. To better manage the clients, we propose a multi-level FL architecture and leverage portfolio theory for incentive mechanism design. Extensive experiment results demonstrate that our approach can obtain better performance when compared to the benchmark methods.

\section{Acknowledgement}
This work is supported in part by Funding for Scientific Research from Jxufe, in part by the Sichuan Provincial Key Research and Development Program (2023YFG0118) and in part by the CCF-NSFOCUS ‘Kunpeng’ Research Fund (No. CCF-NSFOCUS 2023013).

\bibliographystyle{ieeetr} 
\bibliography{references.bib}
\end{document}